\documentclass[12pt]{iopart}
\usepackage{hyperref}
\usepackage{graphicx}
\usepackage{color}
\newcommand{\diff}{\mathrm{d}}

\begin{document}

\title[]{A shadow of the repulsive Rutherford scattering and Hamilton vector}

\author{D A Shatilov$^1$ and Z K Silagadze$^{1,2}$}
\address{$^1$ Novosibirsk State University, Novosibirsk 630090, Russia}
\ead{d.shatilov@g.nsu.ru}
\address{$^2$ Budker Institute of
Nuclear Physics, Novosibirsk 630 090, Russia}
\ead{Z.K.Silagadze@inp.nsk.su}

\begin{abstract}
The fact that repulsive Rutherford scattering casts a paraboloidal shadow
is rarely exploited in introductory mechanics textbooks. Another rarely used 
const\-ruction in such textbooks is the Hamilton vector, a cousin of the more
famous Laplace-Runge-Lenz vector. We will show how the latter (Hamilton's 
vector) can be used to explain and clarify the former (paraboloidal shadow).
\end{abstract}
\maketitle

\section{Introduction}
Recently, in an interesting paper \cite{1}, a shadow of the repulsive 
Rutherford scattering was considered in great detail in fixed-target and 
center-of-mass frames. It was noted that this shadow, in addition to playing 
an important role in low-energy ion scattering spectroscopy, is of great 
educational value. 

Despite its educational attractiveness, this simple fact concerning repulsive 
Cou\-lomb orbits was seemingly forgotten. After the boundary parabola equation 
was derived for the envelope of a family of scattered orbits in \cite{1A}, the 
problem did not attract attention and disappeared into oblivion, as witnessed
by the accidental rediscovery of parabolic  shadow by an undergraduate student 
in computer simulations \cite{1B}.

In this article, we will show how the Hamilton vector can be used to obtain 
the paraboloidal shape of the shadow in a simple and transparent way. We hope
that the publication of \cite{1} will stimulate interest in this neither too 
trivial nor too difficult problem, and help it to finally find its way into 
introductory textbooks on classical mechanics.

\section{Hamilton vector}
The Hamilton vector is an extra constant of motion in the Kepler problem, like 
the well-known Laplace-Runge-Lenz vector (in fact, the two are simply related)
\cite{2}. This ``lost sparkling diamond of introductory level mechanics''
\cite{3} was well known in the past, but then almost forgotten (see \cite{2,
4,5,6} and references therein).

A great virtue of the Hamilton vector is that it can be introduced in a 
simple and natural way \cite{6A}. Let 
\begin{equation}
{\bf e}_r=\cos{\phi}\,{\bf i}+\sin{\phi}\,{\bf j},\;\;
{\bf e}_\phi=-\sin{\phi}\,{\bf i}+\cos{\phi}\,{\bf j},
\label{eq1}
\end{equation}
be time-dependent polar-system unit vectors. Then (the dot denotes 
time-derivative)
\begin{equation}
\dot{\bf e}_r=\dot{\phi}\,{\bf e}_\phi,\;\;
\dot{\bf e}_\phi=-\dot{\phi}\,{\bf e}_r.
\label{eq2}
\end{equation}
Using the second equation in (\ref{eq2}), we can write Newton's second
law for the Cou\-lomb/Kepler problem in the form (in fixed-target frame)
\begin{equation}
 \dot{\bf V}=\frac{\alpha}{\mu r^2}\,\dot{\bf e}_r=-\frac{\alpha}
{\mu r^2\dot{\phi}}\,\dot{\bf e}_\phi=\frac{\diff}{\diff t}\left(-\frac{\alpha}{L_z}\,
{\bf e}_\phi\right ),
\label{eq3}
\end{equation}
where $\alpha=\frac{Z_pZ_te^2}{4\pi\epsilon_0}$, $Z_p$ and $Z_t$ being  the 
projectile and target charges, respectively, in units of the elementary
charge $e$ and ${\bf V}$ is the the target-relative projectile velocity.
At the last step we have used  that $L_z=\mu r^2\dot{\phi}$,
$z$-component of the total angular momentum of the system, 
is conserved in the central field (like any other components of the angular momentum).
It is important to note that, similar to the traditional formulation of the
two-body problem, a reduced mass of the system is introduced
\begin{equation}
\mu = \frac{m_p m_t}{m_p+m_t},
\label{eq4}
\end{equation}
where $m_p$ and $m_t$ are the projectile and target masses 
in the specified order. Equation (\ref{eq3}) indicates that the following 
vector (the Hamilton vector) is conserved in the Coulomb/Kepler field:
\begin{equation}
{\bf H}={\bf V}+\frac{\alpha}{L_z}\,{\bf e}_\phi.
\label{eq5}
\end{equation}

The physical meaning of the Hamilton vector is especially clear in the 
case of an attractive potential, when $\alpha<0$. Then
\begin{equation}
{\bf H}={\bf V}-{\bf V_c},
\label{eq6}
\end{equation}
where ${\bf V_c}=\frac{|\alpha|}{L_z}\,{\bf e}_\phi$ is the velocity that
the particle would have if it moved in a circular orbit around the center of 
the field with the same $L_z$ component of the angular momentum. Therefore, the
Hamilton vector is a kind of velocity defect (deficit) that is conserved
during motion in the Coulomb/Kepler field.

Note that the better known Laplace-Runge-Lenz vector ${\bf A}$ is 
related to the Hamilton vector:
\begin{equation}
{\bf A}={\bf H}\times {\bf L}={\bf V}\times {\bf L}+\alpha\,{\bf e}_r.
\label{eq7}
\end{equation} 

\section{The shape of the shadow}
\begin{figure}[h!]
\centering
\includegraphics[width=0.8\textwidth]{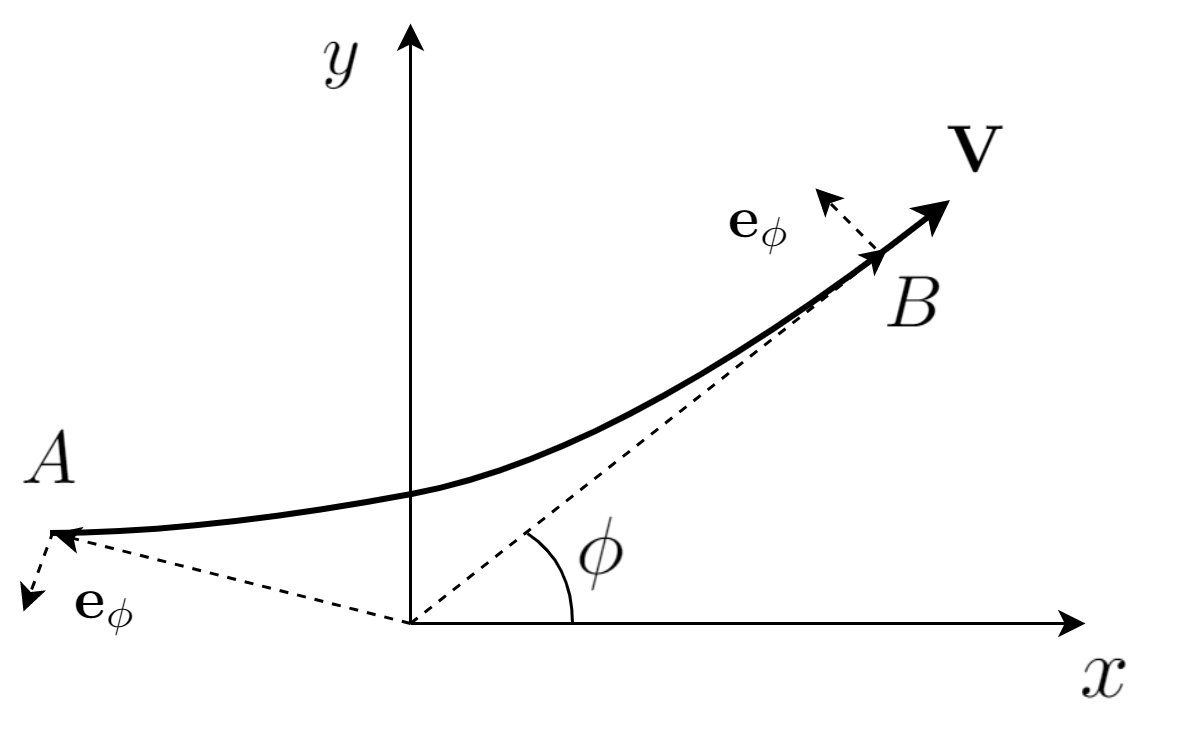}
\caption{Projectile trajectory along which ${\bf H}(A)=
{\bf H}(B)$ for any choice of $A$ and $B$.}
\label{Fig1}
\end{figure}

Let us consider some projectile trajectory (in the fixed-target frame) 
launched at $x=-\infty$ and with an initial velocity $\bf{V_{\infty}}$ 
directed along the x axis. Selected point $B$ at this trajectory has the 
polar angle $\phi$. When point $A$ on this trajectory 
moves towards negative infinity on the $x$-axis, the 
corresponding unit vector ${\bf e}_\phi$ approaches negative direction of the $y$ axis,
and its $x$-projection tends to zero (see figure \ref{Fig1}). 
Let us write the conservation law of the Hamilton vector in the $x$ and
$y$-projections for points $A$ and $B$ in the limit, when the point $A$
corresponds to $x=-\infty$:
\begin{equation}
V_\infty=V_x-\frac{\alpha}{L_z}\sin{\phi},\;\;
-\frac{\alpha}{L_z}=V_y+\frac{\alpha}{L_z}\cos{\phi},
\label{eq8}
\end{equation} 
where ${\bf V}=(V_x,V_y)$ is the projectile velocity at point $B$. Then
\begin{eqnarray} &&
V^2=\left (V_\infty+\frac{\alpha}{L_z}\sin{\phi}\right)^2+\frac{\alpha^2}{L_z^2}
(1+\cos{\phi})^2= \nonumber \\ &&
V_\infty^2+2\,\frac{\alpha^2}{L_z^2}+2\,\frac{\alpha}{L_z}\,
V_\infty\sin{\phi}+2\,\frac{\alpha^2}{L_z^2}\,\cos{\phi}.
\label{eq9}
\end{eqnarray}
On the other hand, $V^2$ can be determined from the energy conservation law
\begin{equation}
\frac{\mu V_\infty^2}{2}=\frac{\mu V^2}{2}+\frac{\alpha}{r},
\label{eq10}
\end{equation}
which gives
\begin{equation}
V^2=V_\infty^2-2\,\frac{\alpha}{\mu r},
\label{eq11}
\end{equation}
where $r$ is the target-relative distance at point $B$. Then, substituting (\ref{eq11}) 
into the l.h.s. of (\ref{eq9}), we get after
a simple algebra
\begin{equation}
\frac{1}{\mu r}=-\alpha\sigma^2(1+\cos{\phi})+\sigma V_\infty \sin{\phi},
\label{eq12}
\end{equation}
where $\sigma=\frac{1}{|L_z|}$ (note that $L_z<0$, because $\dot{\phi}<0$). 
This formula describes every possible (for different initial parameters) 
projectile trajectory, which is a hyperbola in polar coordinates.

The rationale in finding shadow shape from this point can be 
followed as in \cite{1}. Namely, for a given polar angle $\phi$, the shadow 
boundary is determined by the minimum value of $r$. Therefore, we must choose 
$\sigma_m$ (inverse
angular momentum) in such a way that
\begin{equation}
\hspace*{-5mm} 
\left.\frac{\diff}{\diff \sigma}\left [-\alpha\sigma^2(1+\cos{\phi})+\sigma V_\infty \sin{\phi}
\right ]\right|_{\sigma=\sigma_m}=0. 
\label{eq13}
\end{equation}
This gives
\begin{equation}
\sigma_m=\frac{V_\infty\sin{\phi}}{2\alpha(1+\cos{\phi})}.
\label{eq14}
\end{equation}
To calculate the minimum value of $r$ for a given $\phi$, we substitute 
(\ref{eq14}) into  (\ref{eq12}) and get
\begin{equation}
\frac{1}{\mu r_{min}}=\frac{V_\infty^2\sin^2{\phi}}{4\alpha(1+\cos{\phi})}=
\frac{V_\infty^2}{4\alpha} (1-\cos{\phi}).
\label{eq15}
\end{equation}
Therefore, the boundary of the shadow is given by the equation
\begin{equation}
\frac{1}{r_{min}}=\frac{\mu V_\infty^2}{4\alpha} (1-\cos{\phi}),
\label{eq16}
\end{equation}
which is the equation of the parabola in polar coordinates. 

Note that finding the envelope of trajectories from the 
equation (\ref{eq12}) can be put in a general envelope determination context 
by treating this formula as a condition $F(r,\phi,\sigma)=0$
and then applying the general rule for the envelope 
$\frac{\partial}{\partial\sigma} F(r,\phi,\sigma)=0$
(more on this in the concluding remarks).

So far, we have implicitly assumed that all possible projectile 
trajectories are confined to the $x-y$ plane.
In the case of three-dimensional flow, due to the axial symmetry of the 
problem, the shape of the shadow is a paraboloid obtained by rotating the 
parabola (\ref{eq16}) around its axis of symmetry.

Introducing the cylindrical coordinates $z=r_{min}\cos{\phi}$, 
$\rho=r_{min}\sin{\phi}$, equation (\ref{eq16}) can be rewritten in the form
\begin{equation}
\frac{4\alpha}{\mu V_\infty^2}=\sqrt{\rho^2+z^2}-z.
\label{eq17}
\end{equation}
Solving for $ z $, we obtain the following equation for the shadow paraboloid 
in cylindrical coordinates:
\begin{equation}
z=\frac{\rho^2}{8\xi}-2\xi,
\label{eq18}
\end{equation} 
where
\begin{equation}
\xi=\frac{\alpha}{mV_\infty^2}.
\label{eq19}
\end{equation} 

\section{Concluding remarks}
As we can see, using the Hamilton vector allows a simple and transparent way 
to get the main result of \cite{1}, the equation (\ref{eq18}) for the shape 
of the shadow. In fact, we have been using this problem for some time as an
exercise in an introductory mechanics course at Novosibirsk State University 
\cite{7A,7}. In \cite{7}, two more solutions to the problem are given.

The first solution is based on the fact that the shadow boundary is the 
envelope of the projectile trajectories. A one-parameter family of the 
projectile trajectories is given by the equation (for details, see \cite{7}.
Note that in \cite{7} the flow falls from the right).
\begin{equation}
F(r,\phi,\phi_0)=\frac{\alpha}{\mu V_\infty^2}\,\frac{\tan^2{\phi_0}}{r}+1+
\frac{\cos{(\phi-\phi_0)}}{\cos{\phi_0}}=0,
\label{eq20}
\end{equation}   
where $\phi_0$ is the polar angle of the symmetry axis of the
hyperbolic trajectory. Then it is known from mathematics 
(for inquisitive readers: there are some subtleties in the 
mathematical definition of an envelope, see \cite{8,9,10}) 
that the envelope equation is obtained by excluding the parameter $\phi_0$ 
from the system
\begin{eqnarray} &&
F(r,\phi,\phi_0)=0, \nonumber \\ &&
\frac{\partial}{\partial \phi_0} F(r,\phi,\phi_0)=0.
\label{eq21}
\end{eqnarray}
In this way we can obtain the envelope equation
\begin{equation}
\frac{1}{r_{min}}=\frac{\mu V_\infty^2}{4\alpha} (1-\cos{\phi}).
\label{eq22}
\end{equation}
The sign in front of $\cos{\phi}$ can change assuming that the flow falls 
from the right, for example, as it is done in \cite{7}. 
In the present article (as in \cite{1}) it was assumed
that the flow falls from the left.

In the second solution method, considered in \cite{7}, we fix the polar
angle $\phi$ in the family (\ref{eq20}) and choose the parameter $\phi_0$ so
that the corresponding trajectory minimizes $r$. This is the same idea that 
was used in \cite{1} (and in this article), just the parameterizations in the
one-parameter family of projectile trajectories are different.
Therefore, we emphasize that the most original contribution relative to \cite{1} 
is not so much about the procedure for obtaining the shadow itself, but rather 
obtaining trajectories, which is facilitated by the use of the Hamilton vector.

The method used in this article can be extended to the interesting case when
the infalling flux is originated from an emitter at a finite distance from 
the repulsive Coulomb potential centre. Such a modification of the problem of 
the shape of the shadow in repulsive Rutherford scattering is also of 
considerable pedagogical interest. It represents a physical situation that 
actually occurs in atomic physics, and allows students to become familiar 
with such interesting phenomena as the glory and rainbow effects in scattering 
processes \cite{11}.

\begin{figure}[ht!]
\centering
\includegraphics[width=0.8\textwidth]{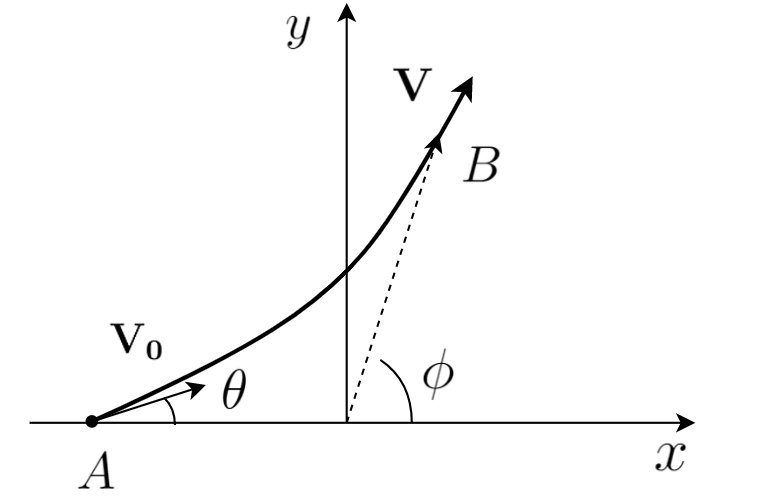}
\caption{The definition of kinematic variables for the projectile trajectory 
launched from the point $A$ on the $x$-axis.}
\label{Fig2}
\end{figure}

Consider projectile trajectories all starting from some point $A$ on the 
$x$-axis at a distance $R$ from the center of the field (see figure \ref{Fig2}). 
We assume that the particles are released from the emitter with the same 
in magnitude initial velocity $V_0$. 
If the initial velocity ${\bf V}_0$ makes an angle $\theta$ with the $x$-axis,
the conservation equations (\ref{eq8}) of the Hamilton vector components will
be modified in the following way:
\begin{equation}
V_{0}\cos{\theta}=V_x-\frac{\alpha}{L_z}\sin{\phi},\;\;
V_{0}\sin{\theta}-\frac{\alpha}{L_z}=V_y+\frac{\alpha}{L_z}\cos{\phi}.
\label{eq23}
\end{equation}
Then
\begin{equation}
V^2={V_0}^2+2 \, \frac{\alpha}{L_z} V_0 \left[\,\sin{(\phi-\theta)} - 
\sin{\theta}\, \right] + 2 \left(\frac{\alpha}{L_z}\right)^2 \left(1 + 
\cos{\phi}\right),
\label{eq24}
\end{equation} 
and using the energy conservation law 
\begin{equation}
\frac{\mu V_0^2}{2}+\frac{\alpha}{R}=\frac{\mu V^2}{2}+\frac{\alpha}{r},
\label{eq25}
\end{equation}
we derive the following relation
\begin{eqnarray} &&
\frac{1}{\mu r}=-\alpha\sigma^2(1+\cos{\phi})+\sigma V_0 \sin{(\phi-\theta)}=
\nonumber \\ &&
-\alpha\sigma^2(1+\cos{\phi})+V_0\sin{\phi}\,\sqrt{\sigma^2-
\frac{1}{\mu^2R^2V_0^2}}-\,\frac{\cos{\phi}}{\mu R},
\label{eq26}
\end{eqnarray}
where we have taken into account that $\frac{1}{\sigma}=|L_z|=\mu V_0R
\sin{\theta}$.

The equation (\ref{eq26}) indicates that the minimum $r$ for a given $\phi$ 
corresponds to $\sigma_m$ such that
\begin{equation}
\sqrt{\sigma_m^2-\frac{1}{\mu^2R^2V_0^2}}=\frac{V_0\sin{\phi}}{2\alpha\,
(1+\cos{\phi})}.
\label{eq27}
\end{equation}
Substituting $\sigma_m$ from (\ref{eq27}) into (\ref{eq26}), we finally obtain
for the shadow boundary the expression
\begin{equation}
\frac{1}{r_{min}}=\frac{\mu V_\infty^2}{4\alpha}(1-A) \left (1-\frac{1+A}{1-A}\,
\cos{\phi}\right ),
\label{eq28}
\end{equation}
where we have introduced a convenient dimensionless parameter \cite{11,12}
\begin{equation}
A=\frac{V_\infty^2}{V_0^2}-1=\frac{2\alpha}{\mu V_0^2R}.
\label{eq29}
\end{equation}
Since $A>0$ equation (\ref{eq28}) corresponds to a hyperbola with  vertex at 
the emission  point  and focus at the force center \cite{11}. As the  emission 
point moves towards infinity, $A\to 0$ and the hyperbola (\ref{eq28}) 
degenerates into the parabola (\ref{eq16}).

For an attractive potential, the envelope of Kepler ellipses, all starting 
from the same point in space, can be obtained in a similar way \cite{12}. 

\section*{Acknowledgments}
The work is supported by the Ministry of Education and Science of the Russian
Federation.

\end{document}